# 2D Monolayer Molybdenum (IV) Telluride TMD: An Efficient Electrocatalyst for Hydrogen Evolution Reaction

**Vikash Kumar[1] and Srimanta Pakhira[1,2*]**

[1] *Theoretical Condensed Matter Physics and Advanced Computational Materials Science Laboratory*, Department of Physics, Indian Institute of Technology Indore, Khandwa Road, Simrol, Indore-453552, MP, India.

[2] *Theoretical Condensed Matter Physics and Advanced Computational Materials Science Laboratory*, Centre for Advanced Electronics (CAE), Indian Institute of Technology Indore, Khandwa Road, Simrol, Indore-453552, MP, India.

*Corresponding author: spakhira@iiti.ac.in (or) spakhirafsu@gmail.com

**Abstract:** An electrocatalyst is needed to efficiently lower the reaction barriers to produce hydrogen through the $H_2$ evolution reaction (HER). Recently, two-dimensional transition metal dichalcogenides (2D TMDs), such as the pure 2D monolayer $MoTe_2$ TMD, have become attractive materials for HER. Using the first principle-based hybrid Density Functional Theory (DFT) method, we have computationally designed a pure 2D monolayer $MoTe_2$ TMD and examined its structural and electronic properties and electrocatalytic efficacy towards HER. A non-periodic finite molecular cluster model $Mo_{10}Te_{21}$ system was employed to explore the feasibility of both the Volmer-Heyrovsky and Volmer-Tafel reaction mechanisms for the HER. The solvent-phase calculations of the HER on the 2D monolayer $MoTe_2$ TMD demonstrate that this material can effectively undergo either Volmer-Heyrovsky or Volmer-Tafel reaction pathways. This conclusion is supported by our determination of low reaction barriers for the H*-migration, Heyrovsky, and Tafel transition states (TSs), which were found to be approximately 9.80, 12.55, and 5.29 kcal.mol$^{-1}$, respectively. These results highlight the potential utility of $MoTe_2$ TMD as a promising electrocatalyst for HER. The unusual electrocatalytic activity of the pure 2D monolayer $MoTe_2$ TMD is evidenced by its ability to significantly reduce reaction barriers, achieving impressive turnover frequency (TOF) values of $3.91 \times 10^3$ and $8.22 \times 10^8$ sec$^{-1}$ during the Heyrovsky and Tafel reaction steps, respectively. Additionally, it demonstrates a remarkably low Tafel slope of 29.58 mV.dec$^{-1}$. These outstanding performance metrics indicate that pure 2D monolayer $MoTe_2$ TMD is a highly efficient electrocatalyst for HER, surpassing the capabilities of traditional platinum group metal-based alternatives. Further exploration of its potential applications in electrocatalysis is warranted. The present work provides valuable insights into the atomic modulation of active sites for enhanced electrocatalytic performance towards HER, paving a way for designing advanced non-noble metal free electrocatalysts.

**TOC:**

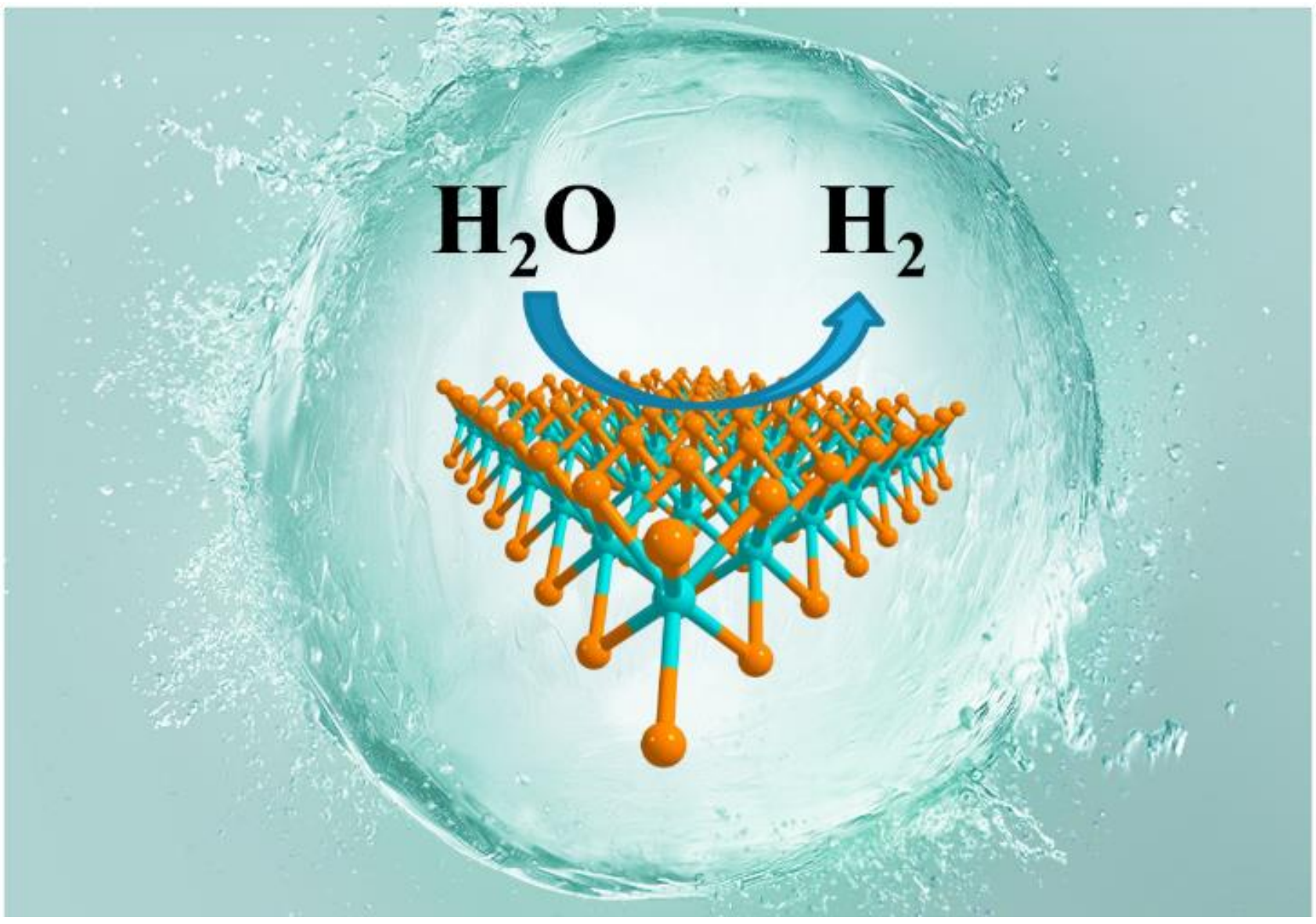

## 1. INTRODUCTION

For centuries, fossil fuels, which include coal, petroleum, natural gas, oil shales, bitumens, tar sands, and heavy oils, have been widely used as a primary source of energy to fulfill the world's energy needs and demands. Due to the widespread use of traditional energy sources like coal, petroleum, natural gas, and oil, there is a growing global energy crisis and environmental pollution, which are both serious issues in today's society and human life.[1] Nonetheless, the extraction and consumption of these conventional energy sources contribute to their depletion and emit harmful pollutants with greenhouse gases to the environment. These emissions are major drivers of climate change, causing severe consequences such as rising global temperatures and detrimental impacts on human health. The pressing concerns surrounding energy depletion and environmental pollution have spurred a growing interest among researchers to explore and develop sustainable and green energy storage and conversion technologies. Among the promising solutions, fuel cells, electrochemical water splitting, and metal-air batteries stand out for their high energy density and eco-friendly features.[2] Hydrogen boasts a substantial energy content and a remarkable energy density when stored in its molecular form, $H_2$. It also occurs abundantly in various organic and inorganic compounds, including water, making it an attractive and efficient candidate for clean energy resources. Most commercially available $H_2$ fuels are produced through a steam reforming process in today's world. During this process, $H_2$ is produced along with the emission of toxic gases such as carbon monoxide (CO) and carbon dioxide ($CO_2$). Water splitting is widely regarded as the most environmentally-friendly approach to produce hydrogen, yielding hydrogen and oxygen as its primary outputs. Among various water-splitting methods, electrocatalytic hydrogen evolution holds a special position in energy storage due to its independence from geographic constraints, weather patterns, and solar intensity fluctuations, making it an exceptionally reliable and efficient technology.[3]

In general, platinum (Pt) and other noble metals are considered to be well-known HER electrocatalysts due to their low overpotentials and small Tafel slopes. Despite its many advantages, the high expense and scarcity of this resource severely curtail its potential applications.[4] Realizing large-scale hydrogen production hinges upon the development of economical and high-performance electrocatalysts that can enhance the energy efficiency of electrochemical water splitting. Developing stable and efficient electrocatalyst materials is essential to achieve hydrogen production on an industrial scale.[5] These materials should be earth-abundant and possess superior properties to facilitate water electrolysis into hydrogen. Extensive research has been conducted to seek alternative materials to replace Pt in hydrogen production. There is a strong emphasis on exploring non-precious metal catalysts that are cost-effective and highly active. These earth-abundant materials are expected to possess exceptional durability and efficiency, making them a viable substitute for precious metal catalysts that are prohibitively expensive.[6,7]

Two-dimensional (2D) materials offer numerous advantages, such as a large specific surface area, exceptional mechanical characteristics, and high carrier mobility. Therefore, in recent years, 2D materials have become one of the most promising candidate materials for electrocatalysts. Recently, 2D layer structure of transition metal dichalcogenides (TMDs) has various special characteristics, including as excellent stability, electrical tunability, high-density active edges, tunable electronic band gap, high electrical conductivity and the potential for defect engineering.[4,8] These exceptional features have garnered significant attention from the scientific community, particularly in the context of studying $H_2$ evolution and its numerous practical applications. In recent times, considerable research has been focused on exploring the potential of 2D transition metal dichalcogenide monolayers (such as $MoSe_2$, $WSe_2$, and $MoS_2$) as substitutes for Pt catalysts. This is mainly due to their remarkable electronic, magnetic, and chemical properties that make them attractive alternatives to traditional catalysts.[6,9] The chemical formula of Earth-abundant TMDs is $MX_2$, where M is a transition metal (TM) atom such as Mo, W, etc., and X is a chalcogen atom such as S, Se, and Te. While the layers of TMDs are weakly held together by out-of-plane van der Waals interactions, the in-plane atoms are held together by strong chemical bonds.[1]

Based on research analysis, it has been discovered that the $MoTe_2$ semiconductor possesses exceptional catalytic activity, particularly in the vicinity of the Fermi energy level, and exhibits high carrier mobility. These properties make $MoTe_2$ a promising candidate for use as an electrocatalyst beyond HER, with potential uses in ORR and OER in fuel cell technology.[10] Here, we introduce 2D monolayer $MoTe_2$ TMD material as a unique and efficient non-noble metal catalyst for the HER. The present research work highlights the various phases of monolayers of molybdenum telluride ($MoTe_2$) as very efficient HER electrocatalysts.[11] A model has been developed to investigate the reaction process and identify the most effective 2D monolayer $MoTe_2$ phase and HER active sites using quantum mechanical density functional theory (DFT) calculations.[5,12–14] This approach offers a comprehensive means of exploring the catalytic activity of the 2D single layer $MoTe_2$, as well as the potential to enhance its performance in comparison to traditional electrocatalytic materials. In this study, we have utilized computational methods to create a 2D $MoTe_2$ TMD monolayer material and investigate its electronic properties, including the electronic band structure, total density of states (DOS), Fermi level ($E_F$) states, and electronic band gap ($E_g$), to gain insight into its potential for HER applications. The density of exposed active edge sites is critical in determining the HER catalytic activity of the 2D monolayer TMDs. The hydrogen adsorption-free energy ($\Delta G_H$) reflects the strength of the interaction between the catalyst and the reactant and is a key indicator of the catalyst's activity. The turnover frequency (TOF) is the rate of hydrogen evolution per active site and is a measure of the catalyst's efficiency. We have utilized hybrid DFT to investigate the HER catalytic activity of the pristine 2D monolayer of $MoTe_2$. Our focus is on determining the reaction pathway of the subject reaction (i.e., HER) on the Mo edge $(10\bar{1}0)$ of the 2D single layer $MoTe_2$ TMD material through theoretical and computational analysis. The exposed Te-edge $(\bar{1}010)$ and Mo-edge $(10\bar{1}0)$ edges of the 2D monolayer TMD $MoTe_2$ have been found to be catalytically active for HER, while the Te-Mo-Te tri-layer of the $MoTe_2$ TMD is the exposed surface. To describe the Mo edge of the 2D monolayer $MoTe_2$, we have used a finite nonperiodic molecular cluster model of $Mo_{10}Te_{21}$ (as shown in Figure 1). The polarization continuum model (PCM) was used to account for the solvation effect of water as a solvent, allowing for the exact integration of DFT to compute reaction barriers.[15–17] The first-principles-based DFT has been used to investigate the reaction pathways, kinetics, and barrier energies for the HER on the surfaces of the 2D monolayer $MoTe_2$ TMD by considering finite cluster model system of the $MoTe_2$ ($Mo_{10}Te_{21}$) TMD as shown in Figure 1.

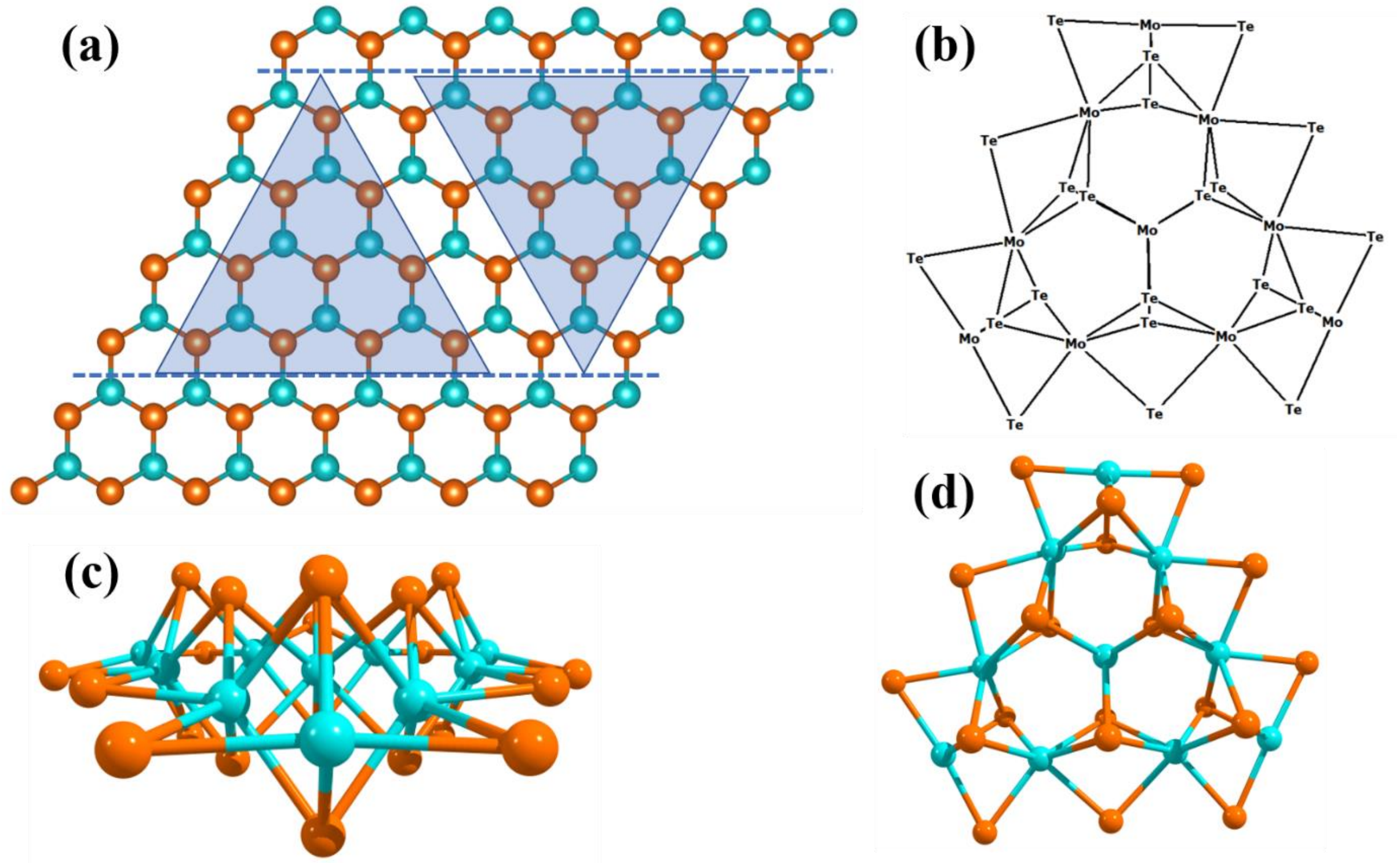

**Figure:1** Figure (a) shows the equilibrium 2D monolayer $MoTe_2$ TMD with a Te-Mo-Te tri-layer structure. The two horizontal blue color dashed lines indicate terminations along the $(10\bar{1}0)$ Mo-edge and $(\bar{1}010)$ Te-edge. The two triangles represent the terminations for Mo-edge and Te-edge clusters. It includes a non-periodic molecular cluster with an active Te-edge and a nonperiodic molecular cluster with an active Mo-edge. (b) The molecular cluster model system of the 2D monolayer $MoTe_2$ TMD is represented schematically. (c) A side view of the chosen $Mo_{10}Te_{21}$ nonperiodic triangular molecular cluster is also shown. (d) A top view of the $MoTe_2$ TMD Mo-edge cluster.

## 2. COMPUTATIONAL DETAILS

Computational methods are used during the present theoretical studies, and the other parameters play a significant role in determining the electrocatalytic activity of an efficient HER. Here, a periodic 2D monolayer structure of the $MoTe_2$ TMD is computationally designed to examine its structural and electronic properties by using the first principle-based hybrid DFT approach.[6,18,19] The HER mechanism at the active edge was further investigated by DFT computations using a nonperiodic finite molecular cluster model system $Mo_{10}Te_{21}$ corresponding to the pristine 2D monolayer $MoTe_2$. A further discussion of both the periodic and nonperiodic systems is explained below.

### 2.1. Periodic Structure DFT Calculations

For the periodic 2D structure calculations, the monolayer of 2D $MoTe_2$ TMD is bounded by Mo-edge ($10\bar{1}0$) and Te-edge ($\bar{1}010$), as shown in Figure 1, and the two horizontal blue color dashed lines indicate terminations along the ($10\bar{1}0$) Mo-edge and ($\bar{1}010$) Te-edge. First, we performed a computational study of 2D monolayer $MoTe_2$ to obtain equilibrium structure and geometry of the TMD. To determine the equilinrium geometry and electronic properties of the 2D monolayer $MoTe_2$ material, we utilized the quatum mechanical first principles based B3LYP-D3 (DFT-D3) method and implemented it in the CRYSTAL17 suite of codes.[20–24] It is expected that the material chemistry of the edges of bulk structure of the $MoTe_2$ may be similar to that of a single $MoTe_2$ layer. It has been shown that additional layers decrease the current density due to electron hopping across the layers, such that the top layers are not as active as the bottom layers. Spin-polarized calculations were also incorporated into the computation by defining the electron occupancy (i.e., α for up spin and β for down spin). The spin-polarized solution was obtained by the keywords "ATOMSPIN" and "SPINLOCK" when the DFT-D method was executed in the CRYSTAL17 program.[13,17] Compared to the Hartree-Fock (HF) method, the DFT method (here, B3LYP-D3) suffers from less or no spin contamination, which helps provide excellent geometry, energy, and electron density calculations. To account for non-bonded weak van der Waals (vdW) interactions between atoms and different layers, we incorporate the semi-empirical Grimme's third-order dispersion correction (Grimme's-D3) in our calculations to obtain accurate equilibrium geometries.[25,26] In order to accurately capture the electronic and structural properties of the system under investigation, we employed the Density Functional Theory with dispersion correction (DFT-D) method, specifically the B3LYP-D3 variant, to account for exchange-correlation effects.[1,27]

During calculations in the CRYSTAL17 program, a Gaussian basis set (GTO) has been used for all atoms. This basis set offers a more efficient alternative to the plane-wave basis set for hybrid DFT calculations. Gaussian basis sets are widely used in quantum chemistry calculations due to their flexibility in accurately describing the electron density around atomic nuclei.[1,2,18,21] They are especially useful for hybrid DFT calculations, which incorporate local and non-local exchange-correlation effects, as they compromise computational cost and accuracy. In comparison, plane-wave basis sets require larger cut-off energies and k-point sampling to achieve similar accuracy in hybrid DFT calculations, which can result in higher computational expenses. In the context of hybrid density functionals, localized Gaussian type of basis set-based codes are better suited for solving the Hartree-Fock (HF) component of the calculation. Gaussian basis sets can handle a wide range of system sizes and molecular geometries, making them a versatile choice for various computational chemistry studies. In this study, we employed triple ζ valence polarized (TZVP) Gaussian type basis sets for both the Mo and Te atoms.[28] We set the value of convergence threshold for evaluating forces, electron density, and energy to $10^{-7}$ atomic units (a.u), ensuring accurate and precise results. Gaussiantype basis sets are particularly well-suited for describing the electron density around atomic nuclei, making them a popular choice for quantum chemistry calculations. To prevent any interactions between the slabs and their periodic images, we implemented a vacuum space of approximately 500 Å in the Z direction of the simulated cells. This value was determined to be adequate based on our current calculations using the CRYSTAL17 suite code.

Using the same level of theory, we conducted electronic properties calculations at the optimized structure of the 2D monolayer $MoTe_2$ TMD material. To accurately capture the electronic properties of the system, we employed Monkhorst k-mesh grids with a size of 20x20x1 to compute the 2D electronic layer structure, geometry, band structure, and total density of states.[29] The use of k-mesh grids allows for a thorough sampling of the Brillouin zone, ensuring accurate results. We calculated eight electronic bands around the Fermi energy level in the high-symmetry *Γ-M-K-Γ* direction within the first Brillouin zone. To accurately capture the electronic states of the 2D monolayer $MoTe_2$ TMD material, we considered all the atomic orbitals of both Mo and Te atoms in calculating the total density of states (DOSs). The electronic band structures and DOS were both estimated with respect to the vacuum in order to account the effects of the electrostatic potential. The equilibrium 2D monolayer structure of $MoTe_2$ TMD was shown using VESTA, a visualization program.[30]

## 2.2 Finite Non- periodic Cluster Modelling

We have computationally developed a $Mo_{10}Te_{21}$, non-periodic finite molecular cluster model system for $MoTe_2$, as shown in Figure 1, to find out the HER mechanism's equilibrium structures, geometries, and reaction barriers. This $Mo_{10}Te_{21}$ cluster model system consists of 10 Mo atoms and 21 Te atoms representing the parent $MoTe_2$ periodic slab structure, as shown in Figure 1. The right triangle represents non-periodic finite cluster models terminating along the Te-edge ($\bar{1}010$), and the inverted triangle corresponds to nonperiodic cluster models terminating along the Mo-edge ($10\bar{1}0$). In the finite molecular cluster model, each Mo atom in the basal plane (001) has a +4-oxidation state and each Mo atom forms six bonds with six adjacent Te atoms. Due to this configuration, a stabilized structure results in each Mo-Te bonding having 4/6=2/3 electron contributions in the inert basal plane. The stabilization of the molecular cluster model also can be understood from the oxidation state of the Te atoms in the basal plane. Each Te atom has -2 oxidation state and creates bonding with 3 Mo atoms, contributing 2/3 electrons towards each Mo-Te bonding in the basal plane. Again, the edges of the molecular cluster model are stabilized with the 2 local electron Mo-Te bonds with a single electron contribution towards four Mo-Te bonds in the basal plane, as shown in Figure 1. This 14/3 {i.e., (2×1) + [4× (2/3)]} contribution of electrons towards the Mo-Te bonds of the edge Mo atom is satisfied with the $d^2$ configuration of one Mo atom and $d^1$ configuration of two Mo atoms at the edges. With this configuration, a stabilized molecular cluster model with periodicity 3 is achieved that derives the molecular cluster model having three edges without any unsatisfied valency. Thus, we considered a molecular cluster $Mo_{10}Te_{21}$ model system (noted by [$MoTe_2$]) to represent the Te-terminated Mo-edges on the surfaces of 2D monolayer $MoTe_2$ TMD as shown in Figure 1, and this $Mo_{10}Te_{21}$ molecular cluster model system is good enough to explain the HER process.[9]

The DFT-M06L method was used to study the HER mechanism on the active surface of $MoTe_2$ through the $Mo_{10}Te_{21}$ finite molecular model. A previously reported report says that the method DFT-M06L gives authentic energy barriers for reaction mechanisms of transition metal-based catalysts.[15,31–33] For all the calculations, we have used 6-31+G** (double-ζ Pople-type) basis sets for O[34] and H[35] atoms and LANL2DZ basis sets for Mo[36] and Te[37] atoms with effective core potentials (ECPs) and include the solvent effect of water by utilizing the polarizable continuum model (PCM).[38] For the PCM calculations, water was taken as a solvent with the dielectric constant of 80.13.[39] PCM is one of the best models to consider the solvation effects, and it is a commonly used method in computational quantum chemistry to model solvation effects. The Zero-Point Vibrational Energy (ZPE) and frequencies were calculated using the same methods and basis sets at the optimized geometry of all the intermediates. Transitions states (TSs) were observed and confirmed by obtaining only one imaginary frequency (negative value) in the modes of vibrations. All the computations were performed with the general-purpose electronic structure quantum chemistry program Gaussian16 to obtain the optimized geometries and TSs to explain the HER mechanism.[40] All the transition state (TS) structures were computed at the optimized geometries and ChemCraft software was used to visualize them.[41] In this study, all the transition states were verified by analyzing the vibrational frequencies and the intrinsic reaction coordinates (IRC).[42,43].

## 3. RESULTS

### 3.1 Structural and electronic properties

The main object of this work is to find the equilibrium 2D monolayer structure of $MoTe_2$ TMD belonging to the 2H phase and find the electrocatalytic actiivitity of the same TMD. The equilibrium 2D monolayer structure of 2H-$MoTe_2$ TMD is shown in Figure 2, obtained by the DFT-D method. By using VESTA software, a 2D monolayer $MoTe_2$ primitive cell was developed, and its structure was optimized by using the first principle-based periodic dispersion-correction hybrid DFT (DFT-D3) method as displayed in Figure 2. The layer slab structure of the 2D monolayer $MoTe_2$ TMD material belongs to the trigonal symetry with a P-6m2 layer group number in the symmetry table. The unit cell of the material is defined by the lattice parameters a = b = 3.40 Å, which is in excellent agreement with previously reported values.[44] Each unit cell consists of one Mo and one Te atom as shown in Figure 2. The equilibrium bond length between Mo and Te atoms was found to be 2.66 Å. The electronic characteristics have been examined by calculating and examining the electronic band structure and densities of states (DOSs) of the 2D single layer $MoTe_2$ TMD,. Analysis of the electronic properties can be useful to obtain an information about electron distribution on the catalytic surface, which is useful for fully understanding the electrocatalytic performance of the TMD.

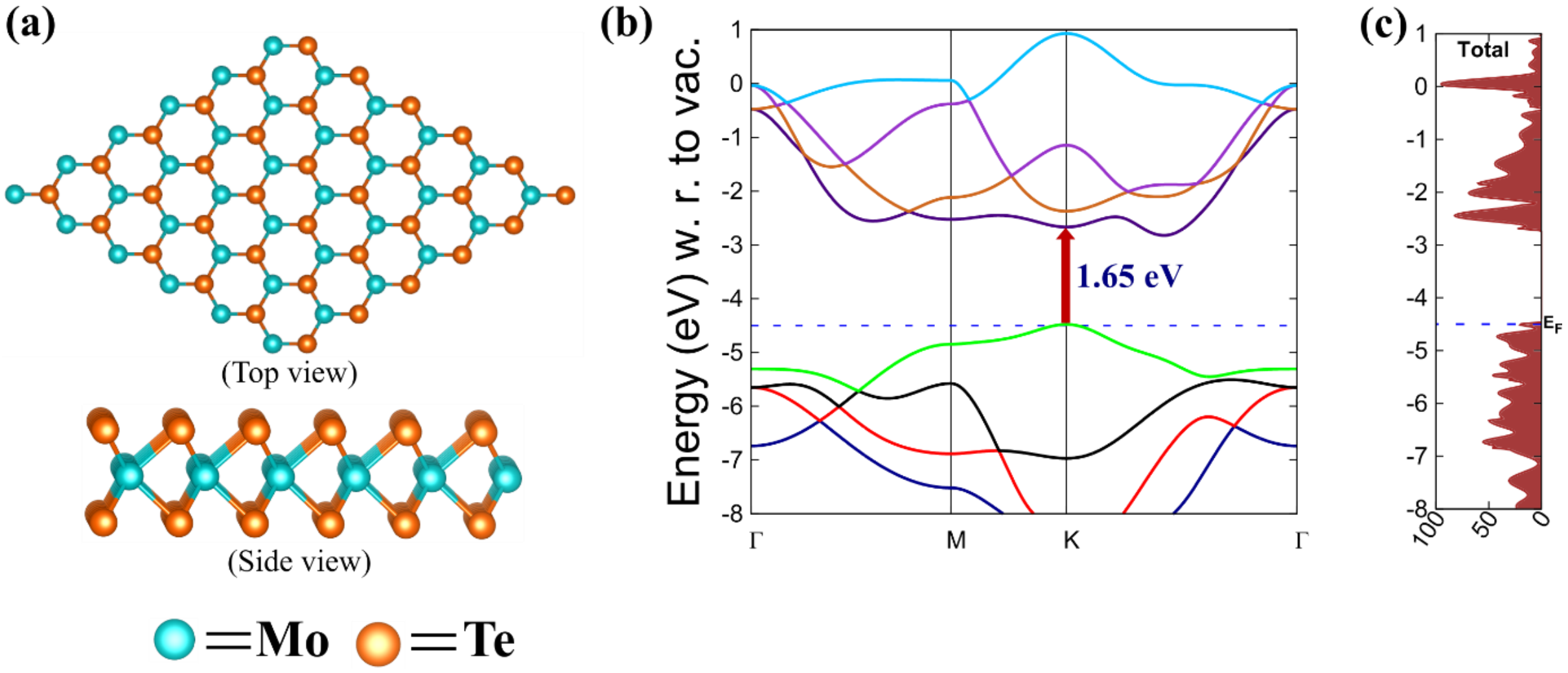

**Figure 2:** Band structure and total density of state (DOS) with the 2D monolayer $MoTe_2$ TMD equilibrium geometry shown here.

We used the DFT-D technique to calculate the equilibrium electronic properties of the 2D single-layer $MoTe_2$ TMDs at equilibrium geometries. The electronic band structure, energy gap ($E_g$), Fermi level ($E_F$), and density of states (DOS) were all computed using the same level of DFT method, were one of the main subjects of our investigation in this present work. This made it possible for us to explore and comprehend the electronic characteristics of the 2D monolayer pristine $MoTe_2$ TMD consistently. We have computed the electronic band structures of 2D monolayer $MoTe_2$ along the *Γ-M-K-Γ* band pathway (which is high symmetric k-path direction consisted with the original symmetry of the 2D slab $MoTe_2$ TMD) with respect to the vacuum, taking into account the initial symmetry of the subject material. The corresponding results are presented in Figure 2b. This approach enabled us to gain valuable insights into the electronic properties of the 2D sinle layer $MoTe_2$ TMD systematically and accurately. In our electronic band structure calculations around the Fermi level ($E_F$) of the 2D monolayer $MoTe_2$, we took into account number of four valence bands (VBs) and four conduction bands (CBs), as depicted in Figure 2b. Our computations revealed that the $E_F$ of $MoTe_2$ was located at -4.50 eV, and the material exhibits a direct band gap of approximately 1.65 eV at the K point, which is much lower than the pristine monolayer $MoS_2$ 2D TMD. Notably, this finding is consistent with earlier reported data, confirming the accuracy and reliability of our calculations.[45] We estimated the density of states (DOS) of the monolayer $MoTe_2$ using the same theoretical framework in order to further study its electronicsl characteristics. The resultant DOS, which corresponds to the electronic band structures of the pure 2D monolayer $MoTe_2$, has been displayed in Figure 2c. The electronic bandgap $E_g$ of the 2D monolayer $MoTe_2$ TMD was determined by our DOS calculations to be around 1.65 eV, which is consistent with the direct band gap found at the K point in the band structure calculations. Thus, it is abundantly clear from our calculations of the electronic properties that the pure 2D monolayer $MoTe_2$ is a semiconductor with a recognizable bandgap which may be useful for electrocatalytic activities towards $H_2$ evolution. The same DFT-D method obtained the intrinsic electronic and structural properties at the equilibrium geometries. After obtaining the equilibrium structure (see Table 1), we computed the electronic properties of the 2D slab structure of the monolayer $MoTe_2$ TMD by employing the B3LYP-D3 method. These findings are crucial for understanding the catalytic performance of the catalyst and finding a suitable electrocatalyst for the hydrogen evolution reaction. The stability of the material has been confirmed by computing the thermodynamic potentials, and the computations are consistent with the previously reported values. The results provide valuable insights into the electronic properties of 2D $MoTe_2$ and its potential applications in electrocatalysis.

**Table 1:** The average equilibrium bond length of the 2D monolayer $MoTe_2$ with optimal equilibrium lattice parameters obtained by the DFT-D method.

| **Materials** | **Lattice parameter (in Å)** | **Bond angles (in °)** | **Bond distance Mo-Se (in Å)** | **Thickness (in Å)** | **Band Gap (in eV)** | **References** |
|---|---|---|---|---|---|---|
| **$MoTe_2$ (Previously reported)** | $a = b =$ 3.51 | $\alpha = \beta =$ 90 $\gamma = 120$ | 2.71 | 3.60 | 1.15 | [44,45] |
| **$MoTe_2$** | $a = b =$ 3.40 | $\alpha = \beta =$ 90 $\gamma = 120$ | 2.66 | 3.59 | 1.65 | This work |

### 3.2 HER Pathway

The hydrogen evolution reaction (HER), which takes place at the cathode of an electrolyzer, is a half-reaction in which protons (in an acidic environment) are reduced, followed by the generation of gaseous hydrogen via the water-splitting process. The overall HER pathway can be described by Eq. (1)[46]

$$H^+ + e^- \rightarrow \frac{1}{2}H_2(g), \qquad \Delta G_H = 0\ eV \qquad (1)$$

Volmer-Heyrovsky or Volmer-Tafel are two possible methods to take place the HER. For the purposes of providing an intermediate state (adsorbed H*) of the processes, it occurs at an electrode in an acidic medium:

i. The Volmer reaction occurs when a proton and an electron combine on the electrode surface to form a hydrogen atom (proton discharge):

$$* + H^+ + e^- \rightarrow *H_{ad} \qquad (2)$$

ii. Electrochemical desorption occurs when the adsorbed hydrogen atom interacts with a proton and an electron, and finally it forms $H_2$. The Heyrovsky reaction is as follows:

$$*H_{ad} + H^+ + e^- \rightarrow H_2 + * \qquad (3)$$

iii. The Tafel reaction results from the coupling of the two hydrogen atoms that have been adsorbed:

$$2 * H_{ad} \rightarrow H_2 + * \qquad (4)$$

The aforementioned fundamental processes result in the Volmer-Heyrovsky and Volmer-Tafel mechanisms. Volmer, Heyrovsky, and Tafel are three rate-determining steps (RDS) that can be used with the aforementioned two methods. Equation 1 can be used to explain the overall HER route during standard conditions. It includes the beginning state, $H^+ + e^-$, the intermediate, adsorbed H*, and the ultimate result, $1/2H_2$ (g). The sum of the energies of $1/2H_2$ (g) and $H^+ + e^-$ is the same. As a result, the change of Gibbs free energy of the intermediate hydrogen adsorption on a catalyst ($\Delta G_H$*), which can be calculated using equation 5, is a crucial indicator of the HER activity of the catalyst.

$$\Delta G_H = \Delta E_H + \Delta E_{ZPE} - T\Delta S_H \qquad (5)$$

where $\Delta E_{ZPE}$ and $\Delta S_H$ stand for the disparity between the zero-point energy of atomic hydrogen adsorption and hydrogen in the gas phase, respectively, and entropy. $\Delta E_H$ is the electronic energy of H upon adsorption. Both $\Delta E_{ZPE}$ and $\Delta S_H$ have negligible and underappreciated catalytic contributions. Equation (5) can finally be condensed into Equation (6):[46]

$$\Delta G_H = \Delta E_H + 0.30\ eV \qquad (6)$$

The adsorption stage will restrict the rate of the overall reaction if the hydrogen-to-surface connection is too weak. The reaction-desorption stage will cap the rate of the total reaction if the hydrogen-to-surface connection is too strong. Hydrogen adsorption energies of ideal HER catalysts are near to $\Delta G_H = 0$, binding hydrogen neither too weak nor too strong.

To study the HER performance of 2D monolayer $MoTe_2$ materials, we created a finite nonperiodic cluster model system of $Mo_{10}Te_{21}$. The HER process is being studied from several angles, including reaction pathways, thermodynamics, chemical kinetics, transition state structures, and reaction barriers. Figures 3 and 6 show the Volmer-Heyrovsky and Volmer-Tafel mechanisms that we looked at in order to assess the HER pathways. We can determine the rate-limiting step by calculating the intermediates and transition states during the production of $H_2$ and by looking at the energy barriers of particular reaction steps. The electrocatalytic properties of the 2D monolayer $MoTe_2$ material have been underexplored, which motivated our study to predict the most dominant HER mechanism for these kinds of TMD materials by analyzing the proposed reaction pathway for both the Vollmer-Heyrovsky and Vollmer-Tafel mechanisms.

### 3.2.1. Volmer-Heyrovsky reaction mechanism

The HER process on 2D $MoTe_2$ TMD material follows the Volmer-Heyrovsky mechanism, which involves a multistep electrode reaction with several intermediates and transition states formed along the reaction pathway. Figure 3 provides a schematic representation of this process, depicting the possible intermediates and transition states formed during the reaction. In this mechanism, the electrocatalyst simultaneously absorbs protons ($H^+$) and electrons ($e^-$) during the Volmer reaction. Then, during the Heyrovsky reaction step, $H_2$ is formed with one proton from the nearby hydronium ion ($H_3O^+$) and H* at the transition metal site in the $MoTe_2$ TMD. To facilitate the whole $H_2$ development process, individual electrons and protons are introduced throughout the HER process. We looked at the stable structures of the nonperiodic finite cluster model of the 2D monoayer $MoTe_2$ TMD ($Mo_{10}Te_{21}$) with sequential additions of each extra electron ($e^-$) and proton ($H^+$) in order to comprehend the fluctuation of free energy between intermediates and identify the pathway with the lowest reaction barrier. The following describes the precise chemical reactions that make up this suggested HER route:

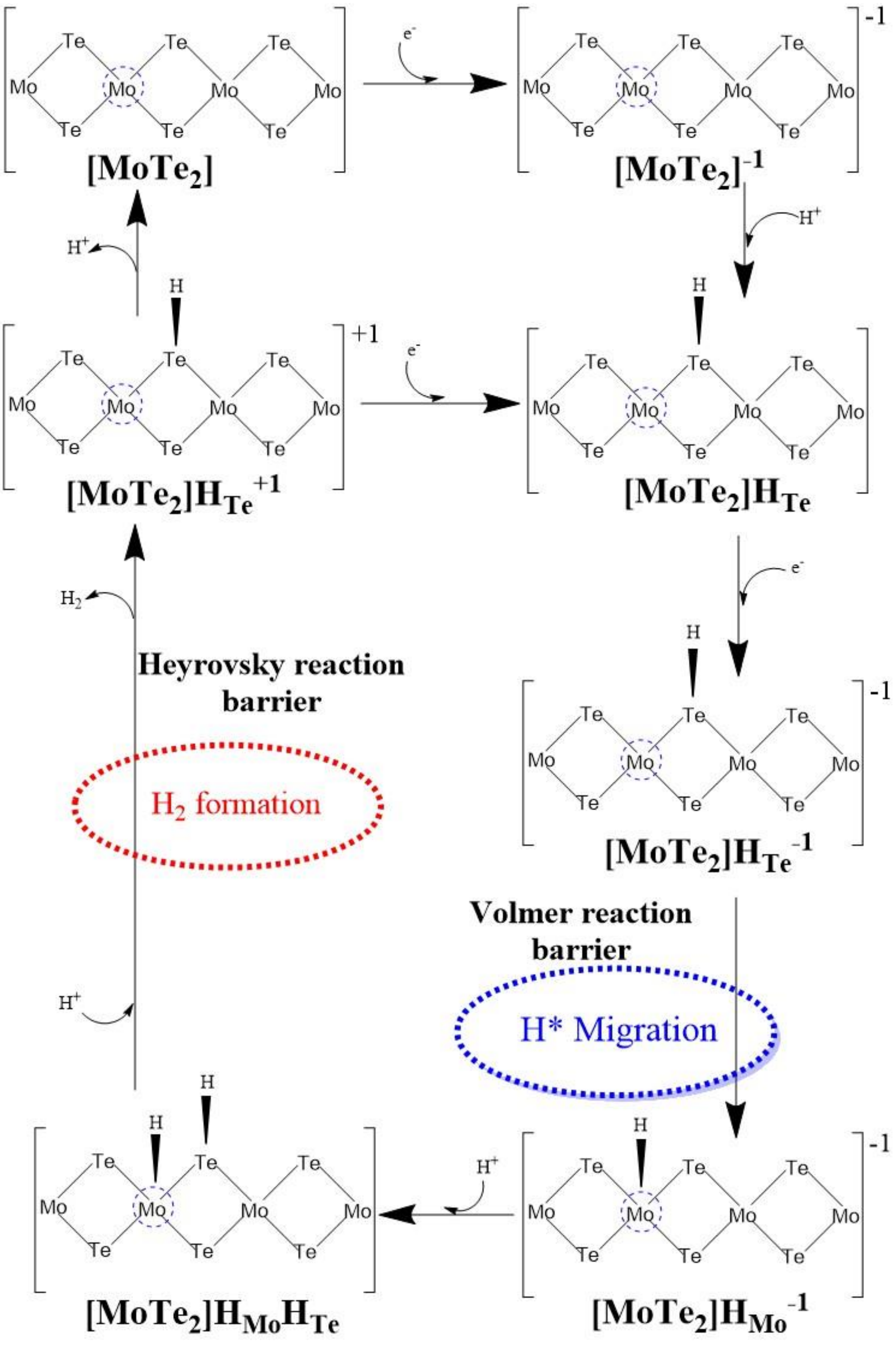

**Figure 3:** Volmer-Heyrovsky HER mechanism the detailed two electron transport reaction pathways of the surface of the 2D $MoTe_2$ material.

1) The [$MoTe_2$] material can be found in its most stable state with a neutral Mo-edge under typical circumstances of the Standard Hydrogen Electrode (SHE) at pH = 0, which serves as the starting point for our estimates of the thermodynamic potential. [$MoTe_2$] represents the abbreviation for the finite molecular cluster $Mo_{10}Te_{21}$. Figure 4a shows equilibrium structure of the [$MoTe_2$] material.
2) The [$MoTe_2$] material absorbs an electron onto its surface to start the HER process, which produces a negatively charged cluster of [$MoTe_2$]$^{-1}$ that is solvated in water and has a delocalized electron on its surface. The DFT technique yielded a value of roughly -424.97 mV for the first reduction potential of the suggested chemical pathway, which entails generating [$MoTe_2$]$^{-1}$ from the pure [$MoTe_2$] by introducing a single electron. Figure 4b shows the equilibrium geometry of [$MoTe_2$]$^{-1}$.
3) According to a study[9], the first hydrogen atom has a strong preference for binding to the Te-edge rather than the Mo atom. As a result, an intermediate molecule called [$MoTe_2$]$H_{Te}$ is created when a proton ($H^+$) is added to the Te-edge, which has an additional electron. The subscript Te denotes the hydrogen atom's bond to the Te atom. The formation of this intermediate incurs an energy cost of approximately 1.93 kcal.mol$^{-1}$. The Te-H equilibrium bond length in the [$MoTe_2$]$H_{Te}$ intermediate's equilibrium structure is 1.67 Å. The equilibrium geometry of the complex is depicted in Figure 4c.
4) A second reduction happens after adding one more electron to the [$MoTe_2$]$H_{Te}$ intermediate, producing [$MoTe_2$]$H_{Te}^{-1}$ (as shown in Figure 4d). The second reduction potential is roughly equal to -766.68 mV.
5) The hydride ion (H*) moves from the Se-site to the neighboring Mo-site in the following process, creating a transition state (TS) of [$MoTe_2$]$H_{Te}^{-1}$. It is the first transition state to arise during the hydrogen evolution reaction (HER) process and is known as the H*-migration reaction or Volmer transition state (TS1). In order to locate the TS, a harmonic vibrational frequency analysis was carried out, and intrinsic reaction coordinate (IRC) calculations were done in order to confirm the existence of TS1.[47,48] During the transfer of H* from the Te-site to the Mo-site, it was discovered that this TS1 had a single hypothetical vibrational frequency, i.e., one imaginary frequency. The equilibrium geometry of TS1 is shown in Figure 4e. It is interesting to note that the current DFT work found that, when computed in the gas phase, the activation energy barrier for the H* migration reaction to create TS1 in the pure 2D monolayer $MoTe_2$ is about G = 8.47 kcal.mol$^{-1}$.

6) Using the DFT approach, it was found that the energy needed to create the $[MoTe_2]H_{Mo}^{-1}$ complex from TS1 was around G = -18.69 kcal.mol$^{-1}$. The equilibrium geometry of the system is shown in Figure 4f. The Mo-H equilibrium bond length in the $[MoTe_2]H_{Mo}^{-1}$ intermediate equilibrium structure is about 1.73 Å. The M06-L DFT method was used to calculate the changes in electronic energy (ΔE), relative enthalpy (ΔH), and Gibbs free energy (ΔG) that occurred during the HER process in each of the several reaction stages and are shown in Table 2.
7) To form the $[MoTe_2]H_{Mo}H_{Te}$ complex, an additional $H^+$ from the solvent medium was added to the Se-site of the $[MoTe_2]H_{Mo}^{-1}$ complex, resulting in a complex where one hydrogen is bound to the Mo atom, and the other hydrogen is bound to the Te atom, as shown in Figure 4g. According to Table 2, the DFT calculation showed that this step cost of energy around -1.97 kcal.mol$^{-1}$. Using the same DFT technique, the equilibrium structure of the $[MoTe_2]H_{Mo}H_{Te}$ complex was found to have equilibrium bond lengths of Mo-H around 1.72 Å and Te-Happroximately 1.67 Å, respectively. The optimized equilibrium structures of all the intermediates and transition states involved in the HER process are displayed in Figure 4.
8) For $H_2$ evolution, the $[MoTe_2]H_{Mo}H_{Te}$ complex can proceed through the Heyrovsky or Tafel reactions. We added a hydronium water cluster ($3H_2O + H_3O^+$) close to the active site of the $Mo_{10}Te_{21}$ nonperiodic molecular cluster to aid the Heyrovsky reaction process. The cluster forms the $[MoTe_2]H_{Mo}H_{Te} + 3H_2O + H_3O^+$ complex as an intermediate with an energy cost of ΔG = -6.08 kcal.mol$^{-1}$ because it has one H at the transition metal Mo site and another H at the Te site. Figure 4h depicts the equilibrium structure of the $[MoTe_2]H_{Mo}H_{Te} + 3H_2O + H_3O^+$ complex as a reaction intermediated formed during the Heyrovsky reaction step.
9) In order to continue with the HER, we formed the second transition state (TS) known as Heyrovsky's transition state (TS2), as depicted in Figure 4i. The $[MoTe_2]H_{Mo}H_{Te}$+$3H_2O$+$H_3O^+$ intermediate, where the H* from the Mo-site and the $H^+$ from the hydronium water cluster combine to produce $H_2$, which separates from the system, is where the Heyrovsky TS2 is obtained. A red circle with a dot in it appears in Figure 4i to represent the generation of $H_2$ during the reaction in TS2. According to the present calculations accomplished in gas phase, the activation energy barrier of the Heyrovsky TS2 is approximately 8.85 kcal.mol$^{-1}$.
10) After the formation of the Heyrovsky TS2, the system changes into $[MoTe_2]H_{Te}^{+1}$, and one $H_2$ molecule and four $H_2O$ molecules are released with an energy cost of -15.34

kcal.mol$^{-1}$. As depicted in Figure 3, this is where the $H_2$ molecule emerges through the catalyst's surface, and the reaction process resumes from the beginning by either absorbing one electron or releasing a proton.

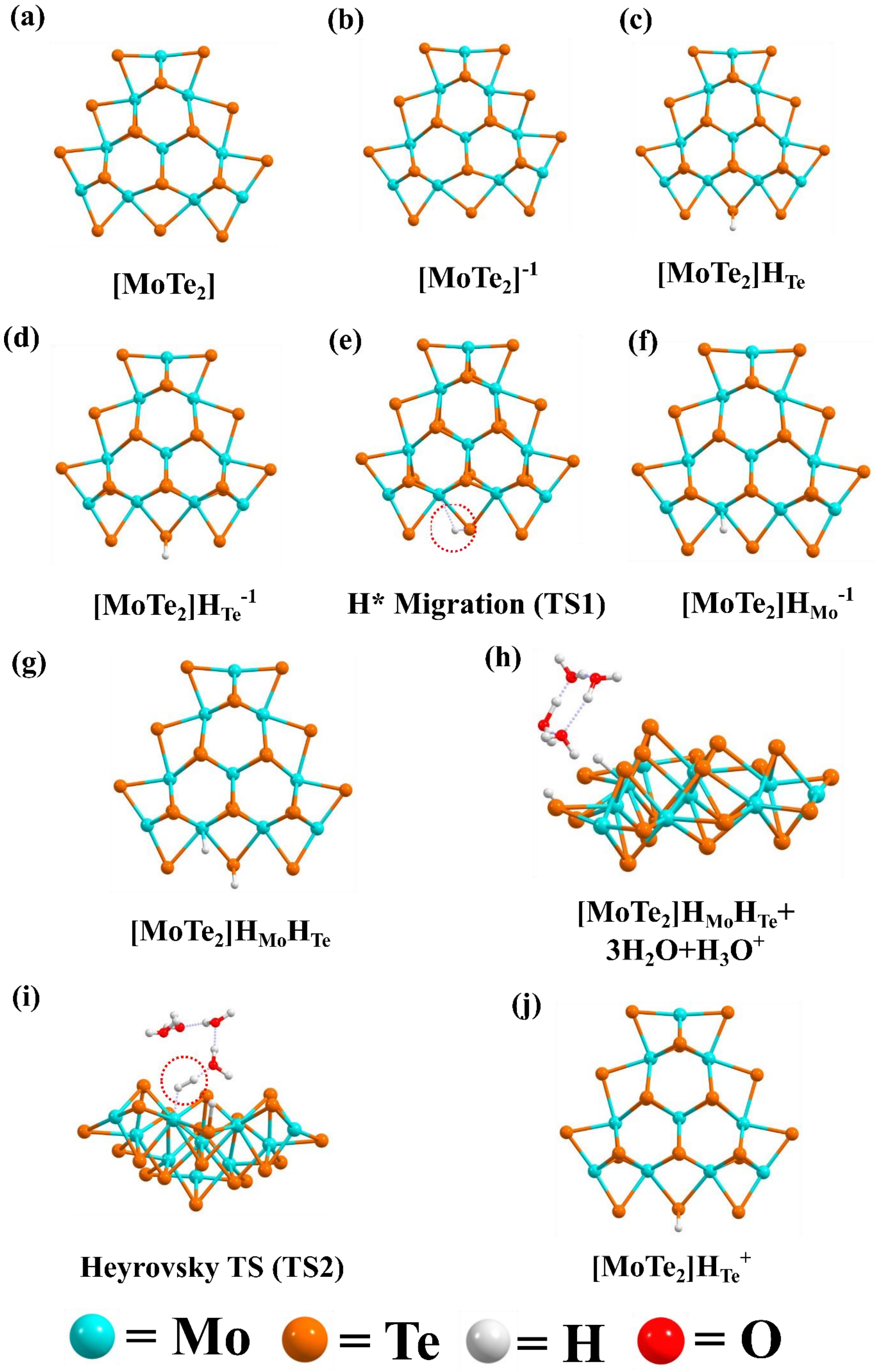

**Figure 4:** Equilibrium geometries of various reaction intermediates with the transition states (TSs) formed during the Volmer-Heyrovsky HER mechanism taken places on the surface of the 2D $MoTe_2$ TMD.

As already mentioned, a very promising route for efficient HER of the catalysts containing transition metals is the Volmer-Heyrovsky mechanism.[9] As a result, we concentrated primarily on determining the activation energy barriers for two important saddle points, the H*-migration TS (noted by TS1) and Heyrovsky TS (noted by TS2). According to gas phase calculations, Table 2 lists the variations in relative electronic energy (ΔE), enthalpy (ΔH), and free energy (ΔG) that occur during the numerous reaction intermediates and TSs involved in the HER via the Volmer-Heyrovsky reaction route.

**Table 2:** Relative electronic energy (ΔE), enthalpy (ΔH), and free energy (ΔG) for various intermediates and transition states (TSs) during HER process followed by the Volmer-Heyrovsky reaction mechanism are computed in the gas phase.

| | **HER Reaction Intermediates** | **ΔE ($kcal.mol^{-1}$) Gas Phase** | **ΔH ($kcal.mol^{-1}$) Gas Phase** | **ΔG ($kcal.mol^{-1}$) Gas Phase** |
|---|---|---|---|---|
| **1** | $[MoTe_2] \longrightarrow [MoTe_2]^{-1}$ | 10.63 | 10.57 | 9.80 |
| **2** | $[MoTe_2]^{-1} \longrightarrow [Nb\text{-}MoTe_2]H_{Te}$ | -2.04 | 2.51 | 1.93 |
| **3** | $[MoTe_2]H_{Te} \longrightarrow [MoTe_2]H_{Te}^{-1}$ | 17.18 | 17.05 | 17.68 |
| **4** | $[MoTe_2]H_{Te}^{-1} \longrightarrow$ Volmer TS1 | 9.50 | 8.18 | 8.47 |
| **5** | Volmer TS $\longrightarrow [MoTe_2]H_{Mo}^{-1}$ | -20.89 | -19.14 | -18.69 |
| **6** | $[MoTe_2]H_{Mo}^{-1} \longrightarrow [MoTe_2]H_{Mo}H_{Te}$ | -6.01 | -1.25 | -1.97 |
| **7** | $[MoTe_2]H_{Mo}H_{Te} \longrightarrow [MoTe_2]H_{Mo}H_{Te}+3H_2O+H_3O^+$ | -21.60 | -20.82 | -6.08 |
| **8** | $[MoTe_2]H_{Mo}H_{Te}+3H_2O+H_3O^+ \longrightarrow$ Herovsky TS2 | 11.13 | 9.32 | 8.85 |
| **9** | Herovsky TS $\longrightarrow [MoTe_2]H_{Te}^{+1}$ | -0.95 | 0.63 | -15.34 |

A reaction barrier of approximately ΔG = 8.47 $kcal.mol^{-1}$ was found in the current investigation for the H*-migration process or TS1 formation, which was obtained in gas phase. Corresponding to this, a reaction barrier of 8.85 $kcal.mol^{-1}$ was obtained (in gas phase calculation) during the TS2 formation in the Heyrovsky reaction process at the Mo-edges of the pure 2D monolayer $MoTe_2$ material. In the case of the pure 2D monolayer $MoTe_2$ TMD, the Heyrovsky reaction step is likely to be the rate-determining step in the Volmer-Heyrovsky reaction mechanism of the HER process given that the Heyrovsky TS2 has slightly higher value of reaction barrier than the H -migration transition state TS1. The HER pathway followed by the Volmer-Heyrovsky reaction mechanism is shown in Figure 5, which shows the fluctuations of relative Gibbs free energies (ΔG) with regard to the reaction coordinates involved in the Volmer-Heyrovsky reaction. The activation energy barriers must also be calculated in the solvent phase because the reactions in commercial fields are typically conducted in solutions. In order to account for the solvent effect of water, the solvent phase calculation was performed using the Polarizaable Continuum Model (PCM) analysis, and the corresponding reaction barriers were noted down, as described in the Volmer-Heyrovsky mechanism steps above.

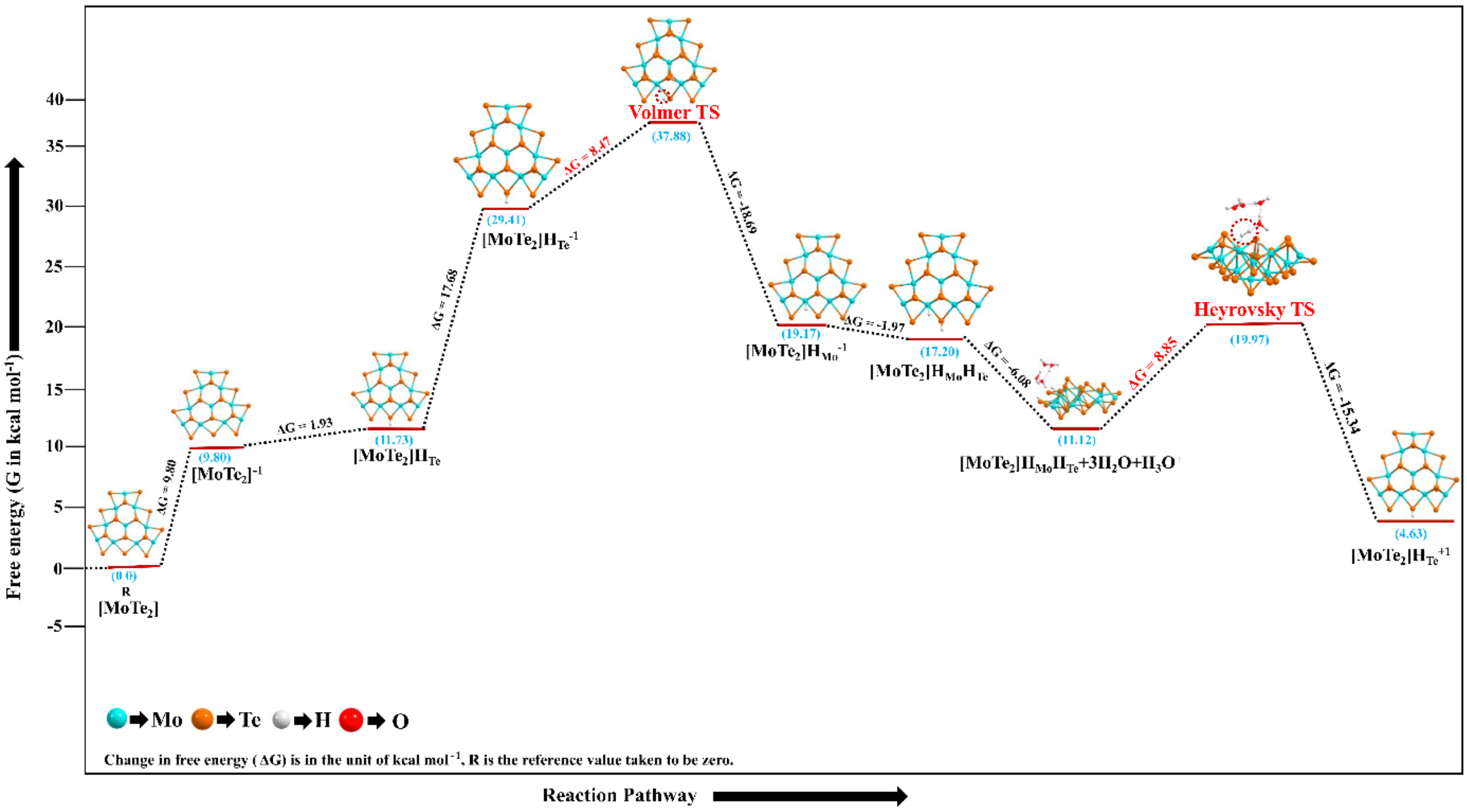

**Figure 5:** The HER pathway followed by the Volmer-Heyrovsky reaction mechanism is shown here as it occurs during the HER process at the surface of the 2D monolayer $MoTe_2$ material.

We used the Polarizaable Continuum Model (PCM) analysis in our study to take account the solvation impact during the HER process. Our DFT-D calculations showed that the energy barrier for the H*-migration during the formation of TS1 in a water environment was about 9.80 kcal.mol$^{-1}$. In comparison to other pristine TMDs, this 2D $MoTe_2$ TMD exhibits better hydrogen migration/adsorption capabilities due to its low energy barrier value in both the solvent and gas phases. We estimated the energy barrier of the TS2 to be about 12.55 kcal.mol$^{-1}$ in the solvent phase during the Heyrovsky reaction step for $H_2$ production and evolution. It is significant to note that this number exceeds the relevant energy barrier determined by calculations in the gas phase. We first investigated the reaction in the gas phase and then carried out solvent phase calculations utilizing the equilibrium geometries of all the reaction intermediates/TSs estimated in the gas phase because in reallity, the HER happens in a solvent phase. The equilibrium geometries of the systems involved in the reaction, which were estimated in the gas phase, were used for the PCM calculations. Usually, the solvent phase has higher reaction barriers than the gas phase. Significant energy changes result from the explicit consideration of the solvent-reactant interaction by the addition of solvent characteristics. Both the gas  and solvent phases are two separate phases with totally different characteristics. Several other kinds of chemical bonds and interactions, including hydrogen bonds, ion-dipole interactions, and weak van der Waal (vdW) forces, are created during the solvation process.

### 3.2.2. Volmer-Tafel reaction mechanism

In comparison to the Volmer-Heyrovsky reaction mechanism, the Volmer-Tafel reaction mechanism is a simpler two-electron transfer process for HER. To create molecular hydrogen, two hydrogen atoms which were adsorbed on the catalyst surface must recombine. Due to the lack of additional solvated protons, this reaction pathway is simpler than the Volmer-Heyrovsky reaction mechanism. According to this mechanism, two neighbouring adsorbed hydrogen atoms on the catalyst surface combine to generate $H_2$ without a solvated proton, as demonstrated by the reaction $H^* + H^* \rightarrow H_2$. The catalyst surface, where this reaction occurs, gets the energy from the room temparute which is required to break through the activation energy barrier by forming TS noted by Tafel TS. Ultimately, the Volmer-Tafel reaction mechanism is a straightforward but significant process that takes place on a catalyst surface. It must be understood and optimized to create effective and long-lasting electrochemical processes. Figure 6 shows the general reaction steps of the Volmer-Tafel reaction mechanism involved in this indicated process.

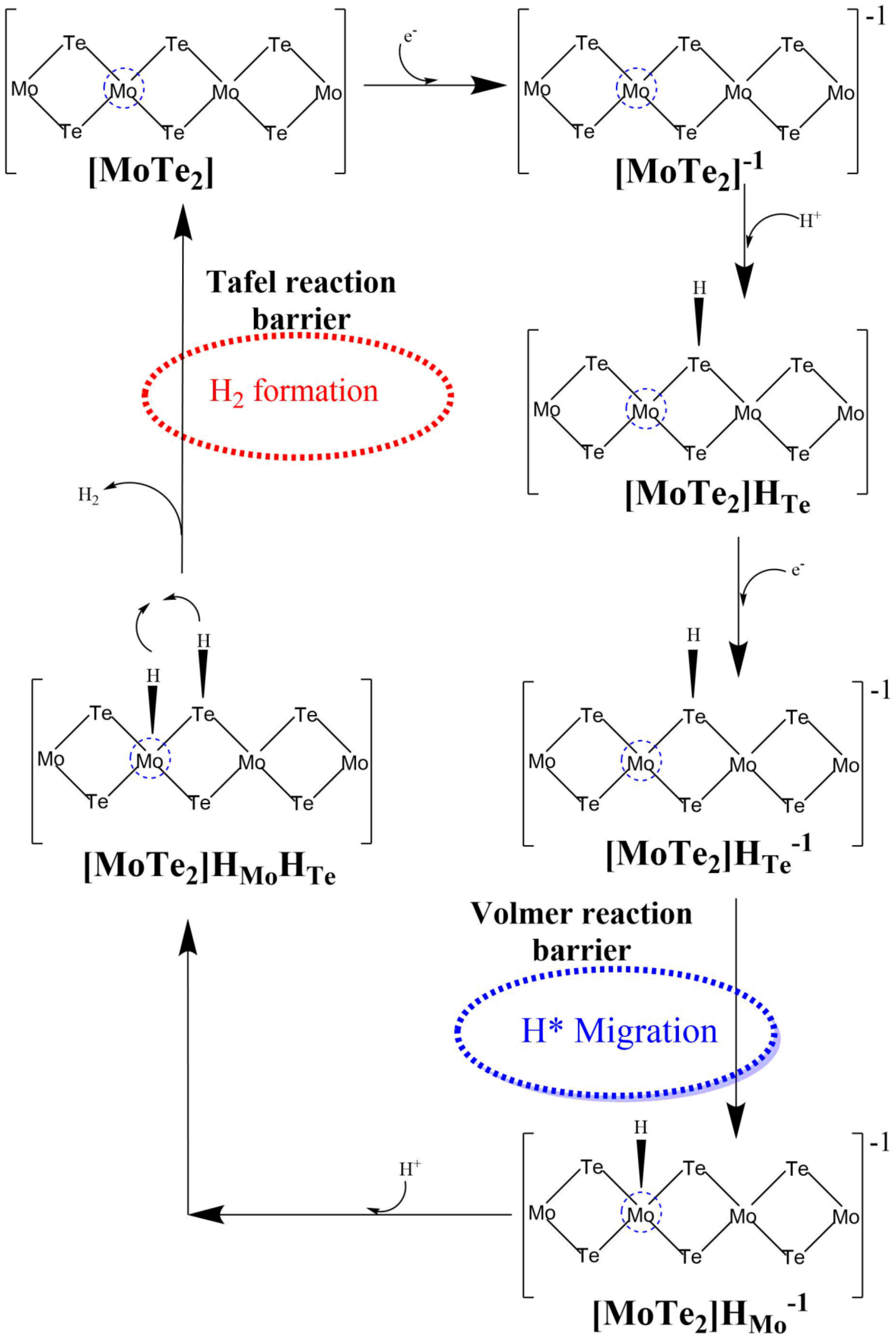

**Figure 6:** Volmer-Tafel HER mechanism with the detailed two electron transfer reaction pathway on the surface of the 2D $MoTe_2$ material.

1) The $[MoTe_2]H_{Mo}^-$ intermediate turns into $[MoTe_2]H_{Mo}H_{Te}$ intermediate by taking an additional proton ($H^+$), and here is the place of $H_2$ evolution where the Volmer-Tafel process diverges from the Volmer-Heyrovsky mechanism. This phase has a -1.97 kcal.mol$^{-1}$ energy cost, which is in consistent with the Volmer-Heyrovsky mechanism pathway. This intermediate, which acts as the starting point for the creation of molecular hydrogen, is essential to the overall reaction mechanism.
2) The interaction of two adsorbed hydrogen atoms, one at the Mo-site and the other at the Te-site, during the HER leads to the formation of $H_2$, which is a result of the Tafel reaction. A transition state (TS3) is created as a result of this reaction, and it has a single imaginary frequency, an equilibrium Mo-H bond length of 1.73 Å, an equilibrium Te-H bond length of 1.68 Å, and a final H-H bond length of roughly 1.74 Å. Using the M06-L DFT approach, it was discovered that the energy barrier of TS3 in the gas phase is 6.07 kcal.mol$^{-1}$. Figure 7 illustrates the TS3's equilibrium geometry and schematic representation. Additionally, using the same level of theory, it was determined that the energy barrier of TS3 for $H_2$ evolution during the Tafel reaction step is approximately 5.29 kcal.mol$^{-1}$ obtained in the solvent phase water. These findings suggest that the energetics and kinetics of the Tafel reaction are strongly influenced by the solvent environment.
3) The last stage of the Volmer-Tafel mechanism, $[MoTe_2]$, is reached once Tafel TS3 is generated. As indicated in Table 3, one $H_2$ molecule is produced from the catalyst's surface in this step, requiring -23.34 kcal.mol$^{-1}$ of free energy. The HER process is finished with the development of $H_2$, giving it an ideal avenue for the creation of clean and sustainable energy.

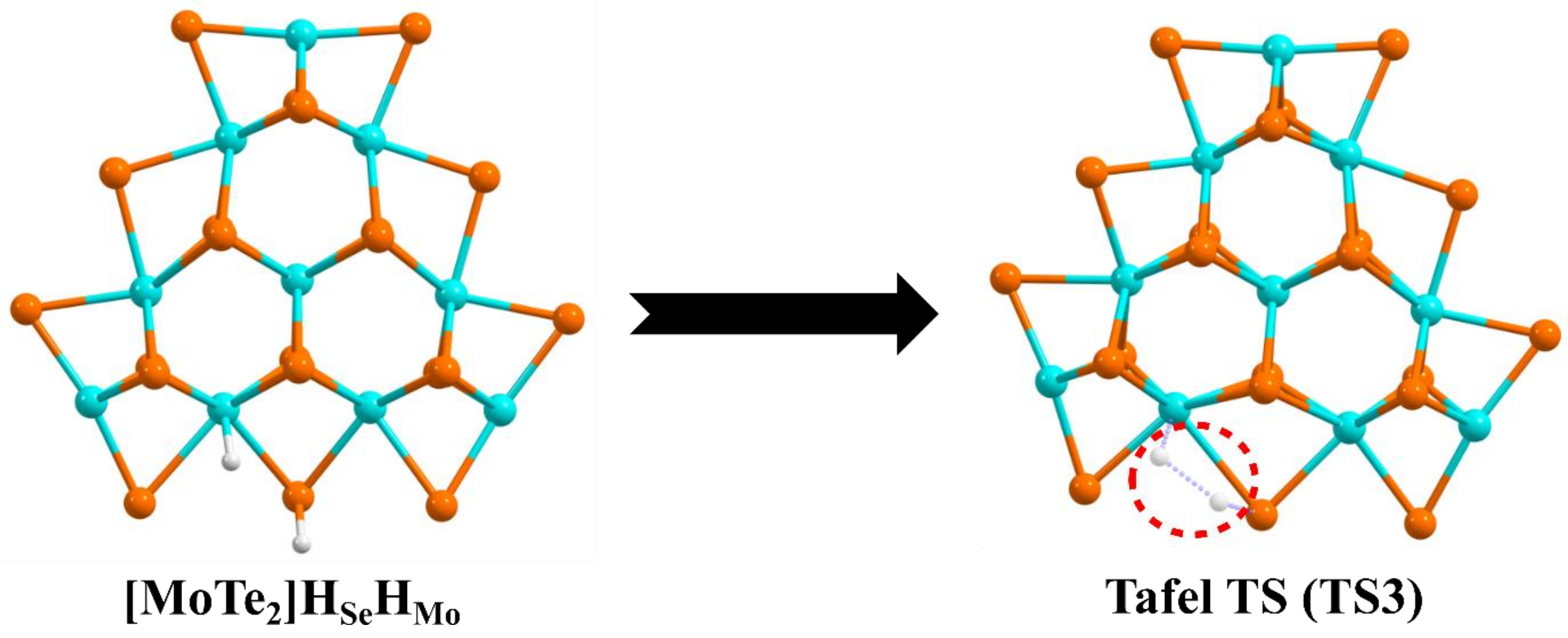

**Figure 7:** The Volmer-Tafel reaction mechanism equilibrium geometries of $[MoTe_2]H_{Mo}H_{Te}$ and TS3 are shown here.

**Table 3:** Below are tabulated energy changes (ΔE, ΔH, and ΔG) for various intermediates and transition states (TSs) during calculations of the Volmer-Tafel reaction mechanism in the gas phase.

| | **HER Reaction Intermediates** | **ΔE ($kcal.mol^{-1}$) Gas Phase** | **ΔH ($kcal.mol^{-1}$) Gas Phase** | **ΔG ($kcal.mol^{-1}$) Gas Phase** |
|---|---|---|---|---|
| **1** | $[MoTe_2] \longrightarrow [MoTe_2]^{-1}$ | 10.63 | 10.57 | 9.80 |
| **2** | $[MoTe_2]^{-1} \longrightarrow [Nb\text{-}MoTe_2]H_{Te}$ | -2.04 | 2.51 | 1.93 |
| **3** | $[MoTe_2]H_{Te} \longrightarrow [MoTe_2]H_{Te}^{-1}$ | 17.18 | 17.05 | 17.68 |
| **4** | $[MoTe_2]H_{Te}^{-1} \longrightarrow$ Volmer TS1 | 9.50 | 8.18 | 8.47 |
| **5** | Volmer TS $\longrightarrow [MoTe_2]H_{Mo}^{-1}$ | -20.89 | -19.14 | -18.69 |
| **6** | $[MoTe_2]H_{Mo}^{-1} \longrightarrow [MoTe_2]H_{Mo}H_{Te}$ | -6.01 | -1.25 | -1.97 |
| **7** | $[MoTe_2]H_{Mo}H_{Te} \longrightarrow$ Tafel TS3 | 6.93 | 5.84 | 6.07 |
| **8** | Tafel TS3 $\longrightarrow [MoTe_2]$ | -14.45 | -14.54 | -23.34 |

In the present calculations for both the gas and the solvent phases, the energy barrier for the Tafel reaction step (TS3) is lower than the energy barrier for the Heyrovsky reaction step (TS2) for the $H_2$ evolution. According to gas phase calculations, the H*-migration step in the TS1 of the Volmer-Tafel reaction pathway of HER has a reaction barrier of around $\Delta G = 8.47$ kcal.mol$^{-1}$, while the Tafel reaction step in the TS3 has a substantially lower reaction barrier of roughly 6.07 kcal.mol$^{-1}$. Figure 8 shows how the Gibbs free energies, also known as HER pathway, diverges relatively with respect to the varius reaction steps in the Volmer-Tafel reaction mechsmism. The values of relative changes in varius energies during the Volmer-Tafel reaction pathway of HER are repored in Table 3.

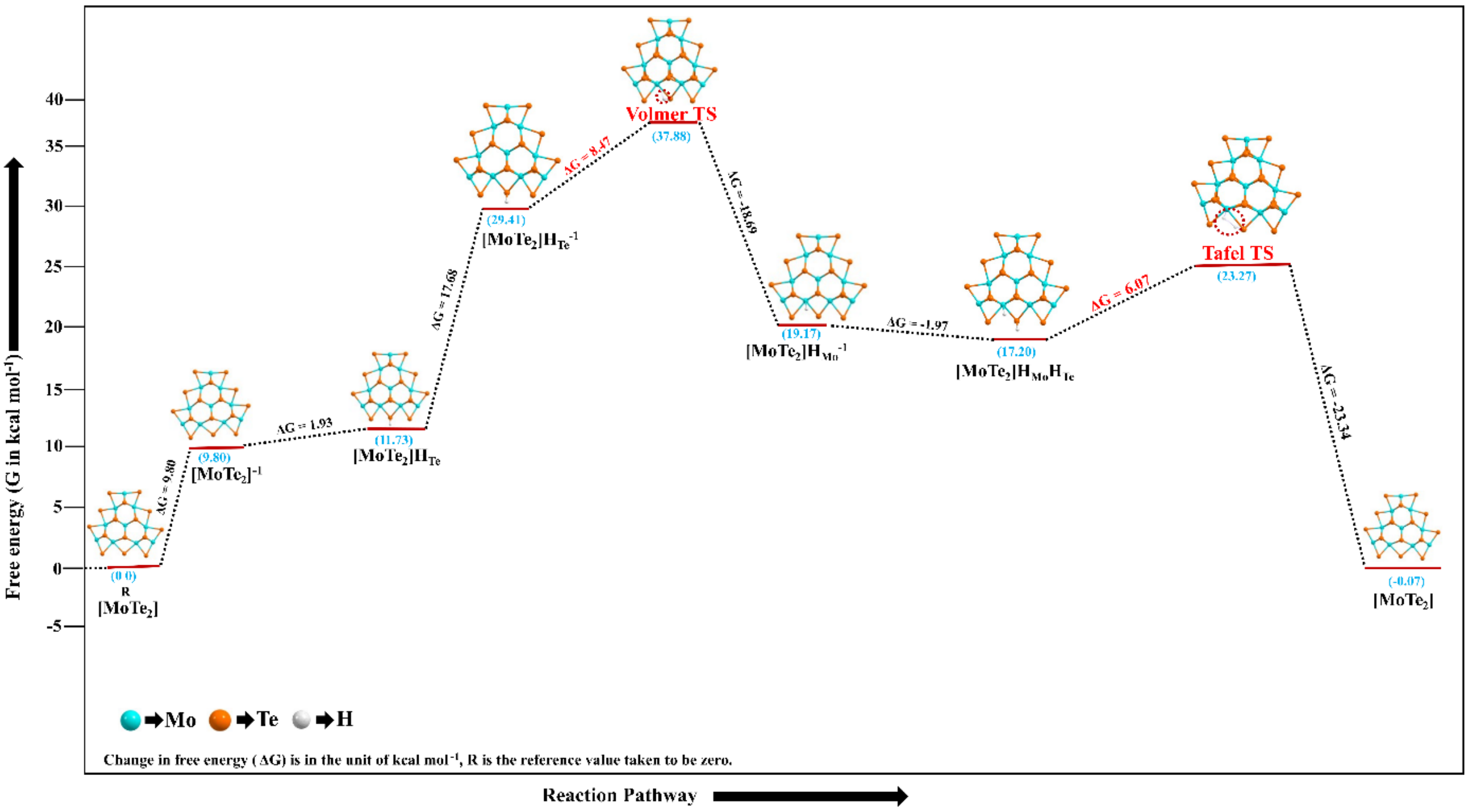

**Figure 8:** The PES of the Volmer-Tafel reaction mechanism is shown here as it occurs during the HER process at the surface of the $MoTe_2$ material.

The $Mo_{10}Te_{21}$ nonperiodic finite molecular cluster model system of the pure 2D monolayer $MoTe_2$ was the subject of this DFT investigation, which looked into the HER on its surfaces. According to the calculations in both the gas and the solvent phases, the activation reaction barrier of the TS3 in the Tafel reaction step ranges from 6.07 to 5.29 kcal.mol$^{-1}$ computed by the DFT method. It is important to note that the reaction energy barrier ($\Delta G$) of the Heyrovsky transition state, TS2, in the Volmer-Heyrovsky reaction mechanism, computed in both the gas and solvent phases, is roughly 2.78 - 7.26 kcal.mol$^{-1}$ higher than the barrier of the TS3, which is the Tafel reaction step. These numbers are within the DFT accuracy range, indicating that both the HER pathways have comparable low energy barriers and can enhance the HER catalytic performance of the 2D monolayer $MoTe_2$ TMD material. With comparable reaction barriers to the noble metals-based catalysts, these decreased reaction barriers suggest that the $H_2$ evolution can occur through either pathway.

**Table 4:** Reaction barriers in both gas and solvent phases for various 2D TMDs.

| Catalysts | H* migration TS1 barrier | | Heyrovsky TS2 barrier | | Tafel TS3 barrier | | References |
|---|---|---|---|---|---|---|---|
| | Gas phase (kcal.mol$^{-1}$) | Solvent phase (kcal.mol$^{-1}$) | Gas phase (kcal.mol$^{-1}$) | Solvent phase (kcal.mol$^{-1}$) | Gas phase (kcal.mol$^{-1}$) | Solvent phase (kcal.mol$^{-1}$) | |
| **Mn-$MoS_2$** | 7.23 | 10.34 | 10.59 | 10.79 | 90.13 | 93.72 | [5] |
| **MoSSe** | 3.93 | 7.10 | 5.61 | 4.72 | 8.52 | - | [3] |
| **$MoTe_2$** | 8.47 | 9.80 | 8.85 | 12.55 | 6.07 | 5.29 | This work |

The current DFT calculations demonstrate that, in comparison to the other materials listed in Table 4, the 2D monolayer $MoTe_2$ material has a lower reaction energy barrier for both the H*-migration TS in the Volmer reaction step and $H_2$ formation in the Heyrovsky and Tafel reaction steps for the HER mechanism. The turnover frequency (TOF) was calculated, and it was discovered that it had a high TOF and a low Tafel slope, as mentioned in the SI. This implies that the 2D monolayer $MoTe_2$ can act as a powerful HER catalyst with high efficiency. We carried out NBO (natural bond orbital), HOMO, and LUMO calculations to determine the TS1, TS2, and TS3 for the H*-migration, Heyrovsky, and Tafel transition states in order to better understand the electrocatalytic activity and stability of this material. According to the calculations, the Tafel TS3, which takes into account the charge cloud and molecule orbital overlap during $H_2$ synthesis, offers a better justification for the effective HER process. We can find more details about the HOMO-LUMO study in the SI.

## 4. CONCLUSIONS

We used first-principles-based DFT-D3 computations (more particularly, the B3LYP-D3 method) to analyze the electronic and structural features of a 2D monolayer slab of the $MoTe_2$ TMD. The present study has found that the 2D $MoTe_2$ monolayer TMD has a direct band gap aboute 1.65 eV. Furthermore, we have studied that the 2D monolayer $MoTe_2$ is a good candidate as an efficient electrocatalyst towards HER because of its lower bandgap and lower activation energy barrier during the HER. We looked at the variation in the adsorption energy of the HER intermediates taken place on the $MoTe_2$ ($Mo_{10}Te_{21}$) surface to examine the catalytic performance of the TMD. Using the M06-L DFT approach, we have carried out each reaction step of both the Volmer-Heyrovsky and Volmer-Tafel reaction pathways to comprehend the HER mechanism route. According to our findings, the Volmer reaction during H*-migration across the 2D $MoTe_2$ TMD material surface from the Te-site to the Mo-site was around 8.47 kcal.mol$^{-1}$ in the gas phase and 9.80 kcal.mol$^{-1}$ in the solvent phase calculations. In the Heyrovsky reaction step, the activation energy barriers for the production of $H_2$ were about 8.85 kcal.mol$^{-1}$ in the gas phase and 12.55 kcal.mol$^{-1}$ in the solvent phase. Also, the activation energy barriers in the Volmer-Tafel HER mechanism were abut 6.07 kcal.mol$^{-1}$ in the gas phase and 5.29 kcal.mol$^{-1}$ in the solvent phase for the production of $H_2$ in the Tafel reaction step. Due to the lower energy barrier of the Tafel TS3 compared to the Heyrovsky TS2, we understood that the Volmer-Heyrovsky mechanism is less advantageous than the Volmer-Tafel mechanism for the HER.

The HOMO and LUMO calculations show that the H*-migration happens during Volmer steps, whereas $H_2$ evolution happens during Heyrovsky or Tafel stages. The extraordinary HER catalytic activity of the 2D monolayer $MoTe_2$ material is explained by this computation. In the solvent phase, the predicted TOF of the Volmer-Heyrovsky reaction mechanism is about $3.91 \times 10^3$ sec$^{-1}$, whereas that of the Volmer-Tafel reaction mechanism is $8.22 \times 10^8$ sec$^{-1}$. The catalyst's efficient hydrogen evolution per active site per unit time is confirmed by these increased TOF values. Also, the theoretical Tafel slope is around 29.58 mV.dec$^{-1}$. This exceptional electrocatalytic activity of the 2D monolayer $MoTe_2$ for HER is demonstrated by the extremely low reaction barrier values, low Tafel slope, and extremely high TOF values. These results imply the need for additional experimental and theoretical research into the potential of 2D monolayer TMD-based materials as HER catalysts.

## Supporting Information

The Supporting Information is available free of charge on the ACS Publications website.

## AUTHOR INFORMATION

**Corresponding Author**

**Dr. Srimanta Pakhira -** *Theoretical Condensed Matter Physics and Advanced Computational Materials Science Laboratory, Department of Physics, Indian Institute of Technology Indore (IITI), Khandwa Road, Simrol, Indore, MP 453552, India.*

*Theoretical Condensed Matter Physics and Advanced Computational Materials Science Laboratory, Centre for Advanced Electronics (CAE), Indian Institute of Technology Indore (IITI), Khandwa Road, Simrol, Indore, MP 453552, India.*

ORCID: orcid.org/0000-0002-2488-300X.

Email: spakhira@iiti.ac.in, spakhirafsu@gmail.com

**Authors**

**Vikash Kumar -** *Department of Physics, Indian Institute of Technology Indore (IIT Indore), Khandwa Road, Simrol, Indore, MP 453552, India.*

ORCID: orcid.org/0000-0002-8811-0583

**Author Contributions**

**Notes**

The authors announce that there are no competing financial interests.

**ACKNOWLEDGEMENT**

The authors are grateful to the Science and Engineering Research Board Department of Science and Technology (SERB-DST), Government of India, under Grant No. ECR/2018/000255 and CRG/2021/000572 for providing financial support for this research work. Dr. Srimanta Pakhira acknowledges the SERB-DST for his Early Career Research Award (ECRA) under project number ECR/2018/000255 and his highly prestigious Ramanujan Faculty Fellowship under scheme number SB/S2/RJN-067/2017. Mr. Vikash Kumar thanks the Indian Institute of Technology Indore (IIT Indore) and UGC, Govt. of India, for his doctoral fellowship UGC Ref. No: 1403/ (CSIR-UGC NET JUNE 2019). The author would like to acknowledge the SERB-DST for providing the computing cluster and programs and IIT Indore for providing the basic infrastructure to conduct this research work. We acknowledge the National Supercomputing Mission (NSM) for providing computing resources of 'PARAM Brahma' at IISER Pune, which is implemented by C-DAC and supported by the Ministry of Electronics and Information Technology (MeitY) and Department of Science and Technology (DST), Government of India. We thank the CSIR, Govt of India for providing the research funds under the scheme no. 22/0883/23/EMR-II and this work is financially supported by the CSIR.